\documentclass[12pt]{iopart}
\usepackage{iopams}
\usepackage{amssymb}
\usepackage{theorem}

\newtheorem{rem}{Remark}

\begin{document}
\title[Mean field Green functions of Hubbard model of superconductivity]
{Rigorous derivation of the mean field Green functions of the
two-band Hubbard model of superconductivity}
\author{Gh Adam and S Adam}
\address{Laboratory of Information Technologies,
        Joint Institute for Nuclear Research,\\
        141980 Dubna, Moscow Region, Russia\\
        and\\
        Horia Hulubei National Institute
        for Physics and Nuclear Engineering
        (IFIN-HH), 407 Atomistilor, Magurele--Bucharest, 077125 Romania}
\ead{adamg@jinr.ru, adams@jinr.ru}
\begin{abstract}
The Green function (GF) equation of motion technique for solving the
effective two-band Hubbard model of high-$T\sb{c}$
superconductivity in cuprates [N.M.~Plakida et al., Phys.~Rev.~B,
{\bf 51}, 16599 (1995); JETP, {\bf 97}, 331 (2003)] rests on the Hubbard
operator (HO) algebra. We show that, if we take into
account the invariance to translations and spin reversal, the HO algebra
results in invariance properties of several specific correlation functions.
The use of these properties allows rigorous derivation and simplification
of the expressions of the frequency matrix (FM) and of the generalized mean
field approximation (GMFA) Green functions (GFs) of the model.
\par
For the normal singlet hopping and anomalous exchange pairing correlation
functions which enter the FM and GMFA-GFs, an approximation procedure based
on the identification and elimination of exponentially small quantities is
described. It secures the reduction of the correlation order to GMFA-GF
expressions.
\end{abstract}

\vspace{2pc}
\submitto{\JPA}
\maketitle
\section{Introduction}
\label{sec:intro}
A consistent theoretical model of the high critical temperature
superconductivity in cuprates is to be able to accommodate both the
normal and superconducting states under incorporation of the essential
features of these systems (see, e.g.,
\cite{Damascelli}
for a review):
strong antiferromagnetic (AFM) superexchange interaction inside the $CuO\sb{2}$
planes, occurrence of two relatively isolated energy bands around the Fermi
level, able to develop $d\sb{x\sp{2}-y\sp{2}}$ pairing: one stemming from
single particle copper $d\sb{x\sp{2}-y\sp{2}}$ states and the second one
from singlet doubly occupied states generated
\cite{ZhR88}
by crystal field interaction;
hopping conduction for an extremely low density of the free charge carriers.
\par
The $p$-$d$ model
\cite{EVar87},
while incorporating all these features, is too cumbersome and cell-cluster
perturbation theory
\cite{FJR96,YOH97}
providing a hierarchy of the various interaction terms was used to derive
simpler models from it. Extreme limit cases of this reduction procedure
are various effective one-band $t$-$J$ models (see, e.g.,
\cite{Pl99,Pl01}
and references therein) which, while unveiling the role played by the
AFM exchange interaction in the occurrence of the $d$-wave pairing,
address exclusively the superconducting state.
\par
The reduction of the $p$-$d$ model to an effective two-band Hubbard model
considered by Plakida et al.~%
\cite{Pl95},
corroborated with the use of the equation of motion technique for
thermodynamic Green functions (GF)
\cite{Zub60},
provided the simplest approach to the description of both the normal
\cite{Pl95,Pl06}
and the superconducting states
\cite{Pl97,Pl03,Pl06a}
within a frame securing rigorous fulfilment of the Pauli exclusion principle
for fermionic states.
\par
The Green function technique rests on the Hubbard operator algebra.
Its rigorous implementation onto a system
characterized by specific symmetry properties (translation invariant
two-dimensional spin lattice, spin reversal invariance of the observables)
results either in characteristic
\emph{invariance properties} of several correlation functions, or in
the occurrence of some \emph{exactly vanishing} correlation functions.
The use of these results allows rigorous derivation and simplification of the
expressions of the frequency matrix and of the generalized mean field
approximation (GMFA) Green functions of the model.
\par
The obtained expressions contain higher order boson-boson correlation
functions (CFs).
For the CFs involving singlets (normal singlet hopping CFs and anomalous
exchange pairing CFs), an approximation procedure which avoids the usual
decoupling schemes and, yet, secures the correlation order reduction to
GMFA-GF expressions, under the identification and
elimination of exponentially small quantities, is described.
\par
The organization of the paper is as follows.
Sec.~\ref{sec:MFA} summarizes essentials of the two-band Hubbard model and
GMFA-GF equations.
Sec.~\ref{sec:transl} describes the invariance properties following from the
translation invariance of the underlying spin lattice.
Sec.~\ref{sec:para} derives invariance properties and
constraints following from the invariance of the macroscopic properties of
the system under spin reversal.
On the basis of the results of Sec.~\ref{sec:transl} and~\ref{sec:para},
rigorous derivation of the frequency matrix in the
(${\bf r}, \omega$)-representation is done in Sec.~\ref{sec:freqmat}.
The derivation of GMFA-GF expressions for the boson-boson correlation
functions involving singlets is discussed in Sec.~\ref{sec:procsing}.
\par
Collecting together the results of sections~\ref{sec:freqmat}
and~\ref{sec:procsing}, expressions of the frequency matrix and of the GMFA
Green function matrix are derived in the $({\bf q}, \omega)$-representation
in sections~\ref{sec:freqmatqo} and \ref{sec:GMFA-GF} respectively.
These results explicitly incorporate both hole-doping and electron-doping
features of the cuprate systems through the singlet hopping and
superconducting pairing terms.
\par
The paper ends with conclusions in section~\ref{sec:concl}.
\section{Mean field approximation}
\label{sec:MFA}
The Hamiltonian of the effective two-band singlet-hole Hubbard model 
\cite{Pl95}
is written in the form
\begin{eqnarray}
    H & = E\sb{1} \sum\sb{i,\sigma} X\sb{i}\sp{\sigma \sigma} +
      E\sb{2} \sum\sb{i} X\sb{i}\sp{22} +
\nonumber\\
& + \mathcal{K}\sb{11}\sum\sb{i,\sigma}\tau\sb{1,i}\sp{\sigma 0,0\sigma} +
   \mathcal{K}\sb{22}\sum\sb{i,\sigma}\tau\sb{1,i}\sp{2\sigma ,\sigma 2} +
   \mathcal{K}\sb{21}\sum\sb{i,\sigma}2\sigma
           (\tau\sb{1,i}\sp{2\bar{\sigma},0\sigma} +
            \tau\sb{1,i}\sp{\sigma 0, \bar{\sigma}2})
\label{eq:H}
\end{eqnarray}
The summation label $i$ runs over the sites of
an infinite two-dimensional (2D) square array the lattice constants
of which, $a\sb{x} = a\sb{y}$, are defined by the underlying single
crystal structure. The spin projection values in the sums over $\sigma$
are $\sigma = \pm 1/2, \bar{\sigma} = -\sigma$.
\par
The Hubbard operators (HOs)
  $X\sb{i}\sp{\alpha \beta} = |i\alpha\rangle \langle i\beta|$
are defined for the four states of the model at each lattice site $i$:
$|0\rangle$ (vacuum), $|\sigma\rangle = \vert \! \! \uparrow \rangle$ and
$|\bar\sigma\rangle = \vert \! \! \downarrow \rangle$ (single particle spin
states inside the hole subband), and
$|2\rangle = \vert \! \! \uparrow \downarrow \rangle$ (singlet state in the
singlet subband).
\par
The multiplication rule holds
  $X\sb{i}\sp{\alpha \beta}X\sb{i}\sp{\gamma\eta} = \delta \sb{\beta\gamma}
   X\sb{i}\sp{\alpha\eta}$.
The HOs may be fermionic (single spin state creation/annihilation in a
subband) or bosonic (singlet creation/annihilation, spin or charge densities,
particle numbers). For a pair of fermionic HOs, the anticommutator rule holds
  $\{X\sb{i}\sp{\alpha \beta}, X\sb{j}\sp{\gamma\eta}\} = \delta\sb{ij}
           (\delta \sb{\beta\gamma}X\sb{i}\sp{\alpha\eta} +
            \delta \sb{\eta\alpha}X\sb{i}\sp{\gamma\beta})$
whereas, if one or both HOs are bosonic, the commutation rule holds
  $[X\sb{i}\sp{\alpha \beta}, X\sb{j}\sp{\gamma\eta}] =
   \delta\sb{ij}(\delta \sb{\beta\gamma}X\sb{i}\sp{\alpha\eta} -
   \delta \sb{\eta\alpha}X\sb{i}\sp{\gamma\beta})$.
At each lattice site $i$, the constraint of no double occupancy of any
quantum state $|i\alpha\rangle$ is rigorously fulfilled due to the
completeness relation
  $X\sb{i}\sp{00} + X\sb{i}\sp{\sigma\sigma} +
   X\sb{i}\sp{\bar{\sigma}\bar{\sigma}} + X\sb{i}\sp{22} = 1$.
\par
In~(\ref{eq:H}), $E\sb{1} = \tilde{\varepsilon \sb{d}} - \mu$ denotes the
hole subband energy for the renormalized energy
$\tilde{\varepsilon \sb{d}}$ of a $d$-hole and the chemical potential $\mu$.
The energy parameter of the singlet subband is
$E\sb{2} = 2E\sb{1} + \Delta$, where
$\Delta \approx \Delta \sb{pd} = \varepsilon \sb{p} - \varepsilon \sb{d}$
is an effective Coulomb energy $U\sb{eff}$ corresponding to the difference
between the two energy levels of the model.
\par
In the description of the hopping processes, the label $1$ points to the
hole subband and $2$ to the singlet subband.
The hopping energy parameter $\mathcal{K}\sb{ab} = 2 t\sb{pd}K\sb{ab}$
depends on $t\sb{pd}$, the hopping $p$-$d$ integral, and on energy band
dependent form factors, $K\sb{ab}$.
Inband ($\mathcal{K}\sb{11}, \mathcal{K}\sb{22}$) and
interband ($\mathcal{K}\sb{21} = \mathcal{K}\sb{12}$) processess are present.
The Hubbard $1$-forms
\begin{equation}
\tau\sb{1,i}\sp{\alpha \beta , \gamma \eta} = \sum\sb{m\neq i}
    \nu\sb{im}X\sb{i}\sp{\alpha \beta} X\sb{m}\sp{\gamma\eta}
\label{eq:taup}
\end{equation}
incorporate the overall effects of specific hopping processes (through the
labels $(\alpha \beta , \gamma \eta)$ of the pair of Hubbard operators)
involving the lattice site~$i$ and its neighbouring sites.
\par
Up to three coordination spheres around the reference site $i$ do contribute
\cite{FJR96,YOH97}
to the sum~(\ref{eq:taup}),
each being characterized by a small specific value of the overlap coefficients
$\nu\sb{ij}$ ($\nu\sb{1}$ for the nearest neighbour (nn), $\nu\sb{2}$ for
the next nearest neighbour (nnn), $\nu\sb{3}$ for the third coordination
spheres).
\par
The quasi-particle spectrum and superconducting pairing for the
Hamiltonian~(\ref{eq:H}) are obtained
\cite{Pl97,Pl03}
from the two-time $4\times 4$ GF matrix (in Zubarev notation
\cite{Zub60})
\begin{equation}
         \tilde G\sb{ij\sigma}(t-t')  =
    \langle\langle \hat X\sb{i\sigma}(t)\! \mid \!
    \hat X\sb{j\sigma}\sp{\dagger}(t')\rangle\rangle =
    -\rmi\theta (t-t')\langle \{\hat X\sb{i\sigma}(t),
    \hat X\sb{j\sigma}\sp{\dagger}\}\rangle ,
\label{eq:GF}
\end{equation}
where $\langle \cdots \rangle$ denotes the statistical average over the
Gibbs grand canonical ensemble.
\par
The GF~(\ref{eq:GF}) is defined for the four-component Nambu column operator
\begin{equation}
  \hat X\sb{i\sigma}=(X\sb{i}\sp{\sigma 2}\,\,
  X\sb{i}\sp{0\bar\sigma}\,\, X\sb{i}\sp{2\bar\sigma}\,\,
  X\sb{i}\sp{\sigma 0})\sp{\top}
\label{eq:nambu}
\end{equation}
where the superscript $\top$ denotes the transposition.
In~(\ref{eq:GF}),
  $\hat X\sb{j\sigma}\sp{\dagger} = (X\sb{j}\sp{2\sigma}\,\,
  X\sb{j}\sp{\bar\sigma 0}\,\, X\sb{j}\sp{\bar\sigma 2}\,\,
  X\sb{j}\sp{0\sigma})$
is the adjoint operator of $\hat X\sb{j\sigma}$.
\par
The GF matrix in (${\bf r}, \omega$)-representation is related to the
expression (\ref{eq:GF}) of the GF matrix in (${\bf r}, t$)-representation by
the non-unitary Fourier transform,
\begin{equation}
    \tilde G\sb{ij\sigma}(t-t')  = \frac{1}{2\pi}
    \int\limits_{-\infty}^{+\infty}\tilde G\sb{ij\sigma}(\omega)
    \; {\rme}\sp{-\rmi\omega (t-t\sp{\prime})}\, {\rmd}\omega \; .
\label{eq:t2om}
\end{equation}
\par
The energy spectrum of the translation invariant spin lattice of (\ref{eq:H})
is solved in the reciprocal space. The GF matrix in this
(${\bf q}, \omega$)-representation is related to the GF matrix in
(${\bf r}, \omega$)-representation by the non-unitary discrete Fourier
transform
\begin{equation}
    \tilde G\sb{ij\sigma}(\omega) =
    \frac{1}{N}\sum\sb{\bf q} {\rme}\sp{-\rmi{\bf q}\;
        ({\bf r}\sb{j} - {\bf r}\sb{i})} \;
    \tilde G\sb{\sigma}({\bf q}, \omega).
\label{eq:r2q}
\end{equation}
\par
For an elemental GF of labels $(\alpha\beta,\gamma\eta)$,
we use the notation
$\langle\langle X\sb{i}\sp{\alpha\beta}(t)|
                X\sb{j}\sp{\gamma\eta}(t')\rangle\rangle$
in the (${\bf r}, t$)-representation
and, similarly,
$\langle\langle X\sb{i}\sp{\alpha\beta}|
                X\sb{j}\sp{\gamma\eta}\rangle\rangle\sb{\omega}$
(assuming Hubbard operators at $t=0$),
in the (${\bf r}, \omega$)-representation.
In the (${\bf q}, \omega$)-representation, it is convenient to use the
notation $G\sp{\alpha\beta,\gamma\eta}({\bf q}, \omega)$.
\par
We shall consider henceforth the GMFA-GF,
$\tilde G\sb{\sigma}\sp{0}({\bf q}, \omega)$.
Its derivation involves:
\par
  (i) Differentiation of the GF~(\ref{eq:GF}) with respect to $t$ and use of
the equations of motion for the Heisenberg operators
$X\sb{i}\sp{\alpha \beta}(t)$.
\par
  (ii) Derivation of an algebraic equation for $\tilde G\sb{ij\sigma}(\omega)$,
Eq.~(\ref{eq:t2om}).
\par
  (iii) Elimination of the contribution of the inelastic processes to the
commutator
$\hat Z\sb{i\sigma}=[\hat X\sb{i\sigma},H]$ entering the equation
of motion of $\tilde G\sb{ij\sigma}(\omega)$.
\par
  (iv) Transformation to (${\bf q}, \omega$)-representation of the obtained
equation of $\tilde G\sp{0}\sb{ij\sigma}(\omega)$ by means of the Fourier
transform~(\ref{eq:r2q}).
\par
This finally yields
\begin{eqnarray}
  \tilde G\sp{0}\sb{\sigma }({\bf q},\omega) = \tilde \chi\;
    \Bigl[ \tilde \chi \omega - \tilde \mathcal{A}\sb{\sigma}({\bf q})
    \Bigr] \sp{-1} \tilde \chi\;,
\label{eq:gf0}\\
   \tilde\chi =
     \langle \{\hat X\sb{i\sigma},\hat X\sb{i\sigma}\sp{\dagger}\}\rangle ,
\label{eq:chi}\\
    \tilde \mathcal{A}\sb{\sigma}({\bf q}) =
    \sum\sb{{\bf r}\sb{ij}} {\rm e}\sp{\rmi{\bf q}\;
        ({\bf r}\sb{j} - {\bf r}\sb{i})} \;
    \tilde \mathcal{A}\sb{ij\sigma}, \quad
      {\bf r}\sb{ij} = {\bf r}\sb{j} - {\bf r}\sb{i}\; ,
\label{eq:Aq2r}\\
  \tilde \mathcal{A}\sb{ij\sigma} = \langle \{ [\hat X\sb{i\sigma}, H],
    \hat X\sb{j\sigma}\sp{\dagger} \} \rangle \, .
\label{eq:Aij}
\end{eqnarray}
The matrix $\tilde \mathcal{A}\sb{ij\sigma}$ is Hermitian.
\section{Translation invariance of the spin lattice}
\label{sec:transl}
Four consequences follow from the translation invariance of the spin lattice.
\begin{itemize}
   \item
      The definition of the Hubbard 1-form~(\ref{eq:taup}) over a
      translation invariant spin lattice results in the identity
      (which secures the hermiticity of the Hamiltonian $H$):
      \begin{equation}
            \tau\sb{1,i}\sp{\alpha \beta, \gamma \eta} =
          - \tau\sb{1,i}\sp{\gamma \eta, \alpha \beta}.
      \label{eq:taudag}
      \end{equation}
   \item
      The Green function~(\ref{eq:GF}) of the model Hamiltonian~(\ref{eq:H})
      depends only on the \emph{distance}
      $r\sb{ij} = |{\bf r}\sb{j} - {\bf r}\sb{i}|$ between the position vectors
      at the lattice sites $i$ and $j$
      \cite{Zub60}.
   \item
      The \emph{one-site} statistical averages are
      \emph{independent} on the site label $i$,
        $\langle X\sb{i}\sp{\alpha\beta}\rangle =
        \langle X\sb{j}\sp{\alpha\beta}\rangle$, ($\forall \ i, j$).
      For this reason, the site label in the one-site averages will be
      omitted.
   \item
      The \emph{two-site} statistical averages
      $\langle X\sb{i}\sp{\alpha\beta} X\sb{j}\sp{\gamma\eta}\rangle$ remain
      invariant under the interchange of the site labels $i$ and $j$,
      \begin{equation}
         \langle X\sb{i}\sp{\alpha\beta} X\sb{j}\sp{\gamma\eta}\rangle =
         \langle X\sb{j}\sp{\alpha\beta} X\sb{i}\sp{\gamma\eta}\rangle ,
         \quad  i\neq j
      \label{eq:av2site}
      \end{equation}
\end{itemize}
\section{Spin reversal invariance}
\label{sec:para}
The energy spectrum of the system described by the Hamiltonian~(\ref{eq:H})
does not depend on the specific values $\sigma = \pm 1/2$ of the spin
projection. As a consequence, the definition of the GF~(\ref{eq:GF}) either
in terms of the $\sigma$-Nambu operator~(\ref{eq:nambu}) or the
$\bar\sigma$-Nambu operator
\begin{equation}
  \hat X\sb{i\bar\sigma}=(X\sb{i}\sp{\bar\sigma 2}\,\,
  X\sb{i}\sp{0\sigma}\,\, X\sb{i}\sp{2\sigma}\,\,
  X\sb{i}\sp{\bar\sigma 0})\sp{\top}
\label{eq:nambusr}
\end{equation}
has to result in mathematically equivalent descriptions of the observables.
This means, however, that the mathematical structures of the frequency matrices
$\tilde \mathcal{A}\sb{ij\sigma}$, Eq.~(\ref{eq:Aij}), and
$\tilde\mathcal{A}\sb{ij\bar\sigma} = \langle \{
[\hat X\sb{i\bar\sigma}, H], \hat X\sb{j\bar\sigma}\sp{\dagger} \} \rangle$
emerging from the $\bar\sigma$-Nambu operator~(\ref{eq:nambusr}), have to be
related to each other.
\par
The identification of the existing relationships is constructive: we
calculate and compare the corresponding matrix elements of
$\tilde \mathcal{A}\sb{ij\sigma}$ and $\tilde \mathcal{A}\sb{ij\bar\sigma}$.
The multiplication rules and the commutation/anticommutation
relations satisfied by the Hubbard operators result in the following general
expression of the elemental anticommutators entering their definitions:
\begin{equation}
    \{[X\sb{i}\sp{\lambda \mu}, H], X\sb{j}\sp{\nu \varphi}\}
        = \delta\sb{ij}C\sb{i}\sp{\lambda \mu, \nu \varphi} +
        (1-\delta\sb{ij})\nu\sb{ij}T\sb{ij}\sp{\lambda \mu, \nu \varphi},
\label{eq:genanticom}
\end{equation}
with \emph{one-site contributions} given by
\begin{eqnarray}
      \fl C\sb{i}\sp{\lambda \mu, \nu \varphi} =
      \delta\sb{\nu \mu}\Big\{ \Big[(\sum\sb{\sigma} \delta \sb{\mu \sigma})
       E\sb{1}+\delta \sb{\mu 2}E\sb{2}\Big]X\sb{i}\sp{\lambda\varphi} +
\nonumber\\
      +\sum\sb{\sigma} \delta \sb{\lambda \sigma}
        \Big[\! -\! E\sb{1}X\sb{i}\sp{\sigma\varphi} \! +\!
         \mathcal{K}\sb{11}\tau\sb{1,i}\sp{0\varphi, \sigma 0} \! -\!
         \mathcal{K}\sb{22}\tau\sb{1,i}\sp{2\varphi, \sigma 2} \! +\!
         \mathcal{K}\sb{21}\!\cdot \! 2\sigma
             (\tau\sb{1,i}\sp{2\varphi,0\bar{\sigma}} \! +\!
              \tau\sb{1,i}\sp{0\varphi,2\bar{\sigma}})\Big] \! +\!
\nonumber\\
         + \delta \sb{\lambda 2}(- E\sb{2}
              X\sb{i}\sp{2\varphi} +
    \mathcal{K}\sb{22}\sum\sb{\sigma}\tau\sb{1,i}\sp{\sigma\varphi,2\sigma} +
  \mathcal{K}\sb{21}\sum\sb{\sigma}2\sigma
                   \tau\sb{1,i}\sp{\bar{\sigma}\varphi,\sigma 0}) -
\nonumber\\
         - \delta \sb{\lambda 0}(
  \mathcal{K}\sb{11}\sum\sb{\sigma}\tau\sb{1,i}\sp{\sigma\varphi,0\sigma} +
  \mathcal{K}\sb{21}\sum\sb{\sigma}2\sigma
                     \tau\sb{1,i}\sp{\sigma\varphi,\bar{\sigma}2})\Big\} +
\nonumber\\
       \! \! \! \! \! \! \! \! \! \! \! \! \! \! \!
       +\delta\sb{\varphi\lambda}\Big\{\! \! -\! \Big[(\sum\sb{\sigma}
       \delta\sb{\lambda\sigma})E\sb{1} +
       \delta\sb{\lambda 2}E\sb{2}\Big] X\sb{i}\sp{\nu\mu} +
\nonumber\\
       + \sum\sb{\sigma} \delta \sb{\mu \sigma}
        \Big[E\sb{1}X\sb{i}\sp{\nu\sigma}\! +\!
         \mathcal{K}\sb{11}\tau\sb{1,i}\sp{\nu 0, 0\sigma}\! -\!
         \mathcal{K}\sb{22}\tau\sb{1,i}\sp{\nu 2, 2\sigma}\! +\!
         \mathcal{K}\sb{21}\cdot 2\sigma
             (\tau\sb{1,i}\sp{\nu 2,\bar{\sigma}0}\! +\!
              \tau\sb{1,i}\sp{\nu 0,\bar{\sigma}2})\Big] +
\nonumber\\
         + \delta \sb{\mu 2}(E\sb{2}X\sb{i}\sp{\nu 2} +
     \mathcal{K}\sb{22}\sum\sb{\sigma}\tau\sb{1,i}\sp{\nu\sigma,\sigma 2} +
     \mathcal{K}\sb{21}\sum\sb{\sigma}2\sigma
                     \tau\sb{1,i}\sp{\nu\bar{\sigma}, 0\sigma}) -
\nonumber\\
         - \delta \sb{\mu 0}
    (\mathcal{K}\sb{11}\sum\sb{\sigma}\tau\sb{1,i}\sp{\nu\sigma,\sigma 0} +
     \mathcal{K}\sb{21}\sum\sb{\sigma}2\sigma
                     \tau\sb{1,i}\sp{\nu\sigma,2\bar{\sigma}})\Big\} -
\nonumber\\
       \! \! \! \! \! \! \! \! \! \! \! \! \! \! \!
     -\sum\sb{\sigma} \delta \sb{\lambda \sigma}\Big[\delta \sb{\varphi 0}
     (\mathcal{K}\sb{11}\tau\sb{1,i}\sp{\nu \mu ,\sigma 0} +
      2\sigma \mathcal{K}\sb{21}\tau\sb{1,i}\sp{\nu\mu, 2\bar{\sigma}}) -
  \delta\sb{\varphi 2}(\mathcal{K}\sb{22}\tau\sb{1,i}\sp{\nu\mu ,\sigma 2} -
     2\sigma \mathcal{K}\sb{21}\tau\sb{1,i}\sp{\nu\mu, 0\bar{\sigma}})\Big] +
\nonumber\\
       \! \! \! \! \! \! \! \! \! \! \! \! \! \! \!
     +\sum\sb{\sigma} \delta \sb{\varphi\sigma}\Big[\delta \sb{\lambda 0}
     (\mathcal{K}\sb{11}\tau\sb{1,i}\sp{\nu \mu ,0\sigma} +
      2\sigma \mathcal{K}\sb{21}\tau\sb{1,i}\sp{\nu\mu, \bar{\sigma}2}) -
   \delta\sb{\lambda 2}(\mathcal{K}\sb{22}\tau\sb{1,i}\sp{\nu \mu , 2\sigma} -
     2\sigma \mathcal{K}\sb{21}\tau\sb{1,i}\sp{\nu \mu, \bar{\sigma}0})\Big] -
\nonumber\\
       \! \! \! \! \! \! \! \! \! \! \! \! \! \! \!
     - \sum\sb{\sigma} \delta \sb{\mu \sigma}\Big[\delta \sb{\nu 0}
     (\mathcal{K}\sb{11}\tau\sb{1,i}\sp{\lambda \varphi , 0\sigma} +
 2\sigma \mathcal{K}\sb{21}\tau\sb{1,i}\sp{\lambda \varphi ,\bar{\sigma}2}) -
 \delta\sb{\nu 2}(\mathcal{K}\sb{22}\tau\sb{1,i}\sp{\lambda\varphi, 2\sigma} -
   2\sigma \mathcal{K}\sb{21}\tau\sb{1,i}\sp{\lambda \varphi , \bar{\sigma}0})
        \Big] +
\nonumber\\
       \! \! \! \! \! \! \! \! \! \! \! \! \! \! \!
     + \sum\sb{\sigma} \delta \sb{\nu \sigma}\Big[\delta \sb{\mu 0}
     (\mathcal{K}\sb{11}\tau\sb{1,i}\sp{\lambda \varphi ,\sigma 0} +
 2\sigma \mathcal{K}\sb{21}\tau\sb{1,i}\sp{\lambda \varphi ,2\bar{\sigma}}) -
 \delta\sb{\mu 2}(\mathcal{K}\sb{22}\tau\sb{1,i}\sp{\lambda\varphi,\sigma 2} -
    2\sigma \mathcal{K}\sb{21}\tau\sb{1,i}\sp{\lambda \varphi , 0\bar{\sigma}})
        \Big]
\nonumber
\end{eqnarray}
and \emph{two-site contributions} given by
\begin{eqnarray}
   \fl T\sb{ij}\sp{\lambda \mu, \nu \varphi} =
     \delta\sb{\nu \mu}\Big[(\sum\sb{\sigma} \delta \sb{\mu \sigma})
       (\mathcal{K}\sb{11}X\sb{i}\sp{\lambda 0}X\sb{j}\sp{0\varphi}\! -\!
        \mathcal{K}\sb{22}X\sb{i}\sp{\lambda 2}X\sb{j}\sp{2\varphi})
   \! +\! (-\delta \sb{\mu 0}\mathcal{K}\sb{11}\! +\!
   \delta \sb{\mu 2}\mathcal{K}\sb{22})
   \sum\sb{\sigma}X\sb{i}\sp{\lambda \sigma}X\sb{j}\sp{\sigma \varphi}\Big]+
\nonumber\\
       \! \! \! \! \! \! \! \! \! \! \! \! \! \! \! \! \!
   + \delta\sb{\varphi\lambda}\Big[(\sum\sb{\sigma}\delta\sb{\lambda\sigma})
       (-\mathcal{K}\sb{11}X\sb{i}\sp{0\mu}X\sb{j}\sp{\nu 0}\! +\!
        \mathcal{K}\sb{22}X\sb{i}\sp{2\mu}X\sb{j}\sp{\nu 2})\! +\!
    (\delta\sb{\lambda 0}\mathcal{K}\sb{11}\! -\!
        \delta\sb{\lambda 2}\mathcal{K}\sb{22})
     \sum\sb{\sigma}X\sb{i}\sp{\sigma \mu}X\sb{j}\sp{\nu \sigma}\Big]\! -\!
\nonumber\\
       \! \! \! \! \! \! \! \! \! \! \! \! \! \! \! \! \!
  \! -\! \sum\sb{\sigma} \delta \sb{\lambda \sigma}\Big\{ \delta \sb{\nu 0}
     \mathcal{K}\sb{11}X\sb{i}\sp{0 \mu}X\sb{j}\sp{\sigma \varphi} -
     \delta \sb{\nu 2}
     \mathcal{K}\sb{22}X\sb{i}\sp{2 \mu}X\sb{j}\sp{\sigma \varphi} +
\nonumber\\
  + \mathcal{K}\sb{21}\! \cdot \! 2\sigma \Big[ \delta \sb{\varphi 0}
      X\sb{i}\sp{2 \mu}X\sb{j}\sp{\nu \bar{\sigma}}\! \! +\! \! 
        \delta\sb{\varphi 2}
      X\sb{i}\sp{0 \mu}X\sb{j}\sp{\nu \bar{\sigma}}\! \! +\! \! 
        \delta\sb{\nu ,-\lambda}
     (X\sb{i}\sp{2 \mu}X\sb{j}\sp{0 \varphi}\! \! +\! \! 
      X\sb{i}\sp{0 \mu}X\sb{j}\sp{2 \varphi})\Big] \Big\} +
\nonumber\\
       \! \! \! \! \! \! \! \! \! \! \! \! \! \! \! \! \!
 \! +\! \sum\sb{\sigma} \delta \sb{\mu \sigma}\Big\{ \delta \sb{\varphi 0}
     \mathcal{K}\sb{11}X\sb{i}\sp{\lambda 0}X\sb{j}\sp{\nu \sigma} -
     \delta\sb{\varphi 2}\mathcal{K}\sb{22}
     X\sb{i}\sp{\lambda 2}X\sb{j}\sp{\nu \sigma}+
\nonumber\\
    + \mathcal{K}\sb{21}\! \cdot \! 2\sigma \Big[ \delta \sb{\nu 0}
      X\sb{i}\sp{\lambda 2}X\sb{j}\sp{\bar{\sigma}\varphi}
    + \delta\sb{\nu 2}X\sb{i}\sp{\lambda 0}X\sb{j}\sp{\bar{\sigma}\varphi}
    + \delta\sb{\varphi ,-\mu}(X\sb{i}\sp{\lambda 2}X\sb{j}\sp{\nu 0}
    + X\sb{i}\sp{\lambda 0}X\sb{j}\sp{\nu 2})\Big] \Big\} +
\nonumber\\
       \! \! \! \! \! \! \! \! \! \! \! \! \! \! \! \! \!
   \! +\! \sum\sb{\sigma} \delta \sb{\nu \sigma}\Big[\delta \sb{\lambda 0}
     \mathcal{K}\sb{11}X\sb{i}\sp{\sigma \mu}X\sb{j}\sp{0 \varphi}\! -\!
     \delta\sb{\lambda 2}
     \mathcal{K}\sb{22}X\sb{i}\sp{\sigma \mu}X\sb{j}\sp{2 \varphi}\! +\!
    \mathcal{K}\sb{21}\! \cdot \! 2\sigma 
(\delta \sb{\mu 0}X\sb{i}\sp{\lambda \bar{\sigma}}X\sb{j}\sp{2 \varphi}\! +\!
 \delta\sb{\mu 2}X\sb{i}\sp{\lambda \bar{\sigma}}X\sb{j}\sp{0 \varphi})
  \Big]\! -
\nonumber\\
       \! \! \! \! \! \! \! \! \! \! \! \! \! \! \! \! \!
   \! -\! \sum\sb{\sigma} \delta \sb{\varphi\sigma}\Big[ \delta \sb{\mu 0}
     \mathcal{K}\sb{11}X\sb{i}\sp{\lambda \sigma}X\sb{j}\sp{\nu 0}\! -\!
     \delta \sb{\mu 2}\mathcal{K}\sb{22}
     X\sb{i}\sp{\lambda \sigma}X\sb{j}\sp{\nu 2}\! +\!
    \mathcal{K}\sb{21}\! \cdot \! 2\sigma
    (\delta \sb{\lambda 0}X\sb{i}\sp{\bar{\sigma}\mu}X\sb{j}\sp{\nu 2}
\! +\! \delta\sb{\lambda 2}X\sb{i}\sp{\bar{\sigma}\mu}X\sb{j}\sp{\nu 0})\Big]+
\nonumber\\
       \! \! \! \! \! \! \! \! \! \! \! \! \! \! \! \! \!
    + \mathcal{K}\sb{21}\! \sum\sb{\sigma} 2\sigma
     (\delta \sb{\lambda 0}\delta\sb{\nu 2}
     X\sb{i}\sp{\sigma \mu}X\sb{j}\sp{\bar{\sigma}\varphi}\!
     -\! \delta \sb{\lambda 2}\delta\sb{\nu 0}
     X\sb{i}\sp{\bar{\sigma}\mu}X\sb{j}\sp{\sigma \varphi}\! -\!
    \delta \sb{\mu 0}\delta \sb{\varphi 2}
     X\sb{i}\sp{\lambda \sigma}X\sb{j}\sp{\nu \bar{\sigma}}
    \! +\! \delta \sb{\mu 2}\delta \sb{\varphi 0}
     X\sb{i}\sp{\lambda \bar{\sigma}}X\sb{j}\sp{\nu \sigma}).
\nonumber
\end{eqnarray}
\par
The comparison of the results obtained from~(\ref{eq:genanticom}) for the
corresponding matrix elements of $\tilde \mathcal{A}\sb{ij\sigma}$ and
   $\tilde \mathcal{A}\sb{ij\bar\sigma}$
and the use of the translation invariance properties~(\ref{eq:taudag})
and~(\ref{eq:av2site}) result in four distinct kinds of relationships:
\begin{itemize}
   \item
     Under the spin reversal $\sigma \rightarrow \bar\sigma$,
     the following \emph{invariance properties} hold for the
     \emph{normal one-site\/} statistical averages:
     \begin{eqnarray}
       && \langle X\sb{i}\sp{\sigma\sigma}\rangle =
          \langle X\sb{i}\sp{\bar\sigma\bar\sigma}\rangle
     \label{eq:avss} \\
       &&\langle \tau\sb{1,i}\sp{\sigma 2, 2\sigma}\rangle =
         \langle \tau\sb{1,i}\sp{\bar\sigma 2, 2\bar\sigma}\rangle , \quad
         \langle \tau\sb{1,i}\sp{0\bar\sigma, \bar\sigma 0}\rangle =
         \langle \tau\sb{1,i}\sp{0\sigma, \sigma 0}\rangle
     \label{eq:avninb} \\
       && 2\sigma \langle \tau\sb{1,i}\sp{\sigma 2, \bar\sigma 0}\rangle =
          2\bar\sigma \langle \tau\sb{1,i}\sp{\bar\sigma 2, \sigma 0}\rangle
     \label{eq:avnitb}
     \end{eqnarray}
   \item
     The identity $\langle C\sb{i}\sp{\sigma 2, 0\sigma} +
     C\sb{i}\sp{0\bar\sigma , \bar\sigma 2}\rangle = 0$ holds,
     therefrom we get for the \emph{one-site anomalous\/} averages,
     \begin{eqnarray}
       &&\langle X\sb{i}\sp{02}\rangle = 0
     \label{eq:av02} \\
       &&\langle \tau\sb{1,i}\sp{0\bar\sigma, \bar\sigma 2}\rangle =
         - \langle \tau\sb{1,i}\sp{0\sigma, \sigma 2}\rangle
     \label{eq:avitb} \\
       &&\langle \tau\sb{1,i}\sp{0\bar\sigma, 0\sigma}\rangle =
         \langle \tau\sb{1,i}\sp{\sigma 2, \bar\sigma 2}\rangle 
     \label{eq:avinb}
     \end{eqnarray}
     \par
     The first two equations imply that the contributions
     of the one-site terms $\langle X\sb{i}\sp{02}\rangle$ and
     $\sum\sb{\sigma}\langle\tau\sb{1,i}\sp{0\bar\sigma,\bar\sigma 2}\rangle$
     to the superconducting pairing \emph{vanish identically irrespective
     of the model details} (like, e.g., the relationship between
     the lattice constants $a\sb{x}$ and $a\sb{y}$).
     \par
     For a rectangular spin lattice ($a\sb{x} \neq a\sb{y}$),
     Eq.~(\ref{eq:avinb}) points to the occurrence of a small
     non-vanishing one-site contribution to the superconducting pairing
     originating \emph{equally\/} in both energy subbands.
     However, over the square spin lattice~(\ref{eq:H}) ($a\sb{x} = a\sb{y}$),
     each term of~(\ref{eq:avinb}) vanishes for $d$-wave pairing due to
     the symmetry in the reciprocal space
     \cite{Pl03}.
   \item
     Under the spin reversal $\sigma \rightarrow \bar\sigma$,
     the following \emph{invariance properties} hold for the
     \emph{two-site\/} statistical averages:
     \begin{eqnarray}
       && \langle X\sb{i}\sp{\sigma\sigma}
                  X\sb{j}\sp{\sigma\sigma}\rangle =
          \langle X\sb{i}\sp{\bar\sigma\bar\sigma}
                  X\sb{j}\sp{\bar\sigma\bar\sigma}\rangle , \quad
          \langle X\sb{i}\sp{\sigma\sigma}
                  X\sb{j}\sp{\bar\sigma\bar\sigma}\rangle =
          \langle X\sb{i}\sp{\bar\sigma\bar\sigma}
                  X\sb{j}\sp{\sigma\sigma}\rangle
     \label{eq:ss} \\
       && \langle X\sb{i}\sp{22}
                  X\sb{j}\sp{\sigma\sigma}\rangle =
          \langle X\sb{i}\sp{22}
                  X\sb{j}\sp{\bar\sigma\bar\sigma}\rangle , \quad
          \langle X\sb{i}\sp{00}
                  X\sb{j}\sp{\sigma\sigma}\rangle =
          \langle X\sb{i}\sp{00}
                  X\sb{j}\sp{\bar\sigma\bar\sigma}\rangle
     \label{eq:2s0s} \\
       && \langle X\sb{i}\sp{02}
                  X\sb{j}\sp{\sigma\sigma}\rangle =
          \langle X\sb{i}\sp{02}
                  X\sb{j}\sp{\bar\sigma\bar\sigma}\rangle .
     \label{eq:02s}
     \end{eqnarray}
   \item
     The operator of the number of particles at site $i$ within the
     singlet subband, $N\sb{i}$, is the sum of spin $\sigma$ and $\bar\sigma$
     components,
     \begin{equation}
       N\sb{i} = n\sb{i\sigma} + n\sb{i\bar\sigma}, \quad
       n\sb{i\sigma} = X\sb{i}\sp{\bar\sigma\bar\sigma}+X\sb{i}\sp{22}, \quad
       n\sb{i\bar\sigma} = X\sb{i}\sp{\sigma\sigma} + X\sb{i}\sp{22}.
     \label{eq:opNi}
     \end{equation}
     Similar relationships hold for the number of particles at site $i$
     within the hole subband, $N\sb{i}\sp{h}$,
     \begin{equation}
       N\sb{i}\sp{h} = n\sb{i\sigma}\sp{h} + n\sb{i\bar\sigma}\sp{h}, \quad
       n\sb{i\sigma}\sp{h} = X\sb{i}\sp{\sigma\sigma} + X\sb{i}\sp{00}, \quad
       n\sb{i\bar\sigma}\sp{h} = X\sb{i}\sp{\bar\sigma\bar\sigma} +
                                 X\sb{i}\sp{00}.
     \label{eq:opNih}
     \end{equation}
     Due to the completeness relation,
     \begin{equation}
       N\sb{i} + N\sb{i}\sp{h} = 2 ,\quad
       n\sb{i\sigma} + n\sb{i\sigma}\sp{h} =
       n\sb{i\bar\sigma} + n\sb{i\bar\sigma}\sp{h} = 1.
     \label{eq:opNisum}
     \end{equation}
     These equalities simply reflect the fact that, at a given lattice
     site $i$, there is a single spin state of predefined spin projection,
     whereas the total number of spin states equals two.
     \par
     Therefore, the operator $N\sb{i}$, Eq.~(\ref{eq:opNi}), provides unique
     characterization of the occupied states within the model
     \cite{Pl95,Pl03,Pl06}.
\end{itemize}
\section{Frequency matrix in (${\bf r}, \omega$)-representation}
\label{sec:freqmat}
A straightforward consequence of 
the results established in section~\ref{sec:para} is the
simplest general expression of the frequency matrix
$\tilde \mathcal{A}\sb{ij\sigma}$, Eq.~(\ref{eq:Aij}):
\begin{equation}
  \tilde {\cal A}\sb{ij\sigma} = \delta\sb{ij} \left(
    \begin{array}{cc}
      \hat{c}\sb{\sigma} & \hat{0}\\
      \hat{0} &
         -(\hat{c}\sb{\bar\sigma})\sp{\top}
    \end{array} \right) + (1 - \delta\sb{ij})\left(
    \begin{array}{cc}
      \hat{D}\sb{ij\sigma} & \hat{\Delta}\sb{ij\sigma}\\
      (\hat{\Delta}\sb{ij\sigma})\sp{\dagger} &
         -(\hat{D}\sb{ij\bar\sigma})\sp{\top}
    \end{array} \right).
\label{eq:Aijs}
\end{equation}
The one-site $2\times2$ matrix $\hat{c}\sb{\sigma}$ is
\emph{Hermitian}, its elements do not depend on the particular lattice
site $i$,
\begin{equation}
  \hat{c}\sb{\sigma} = \left(
    \begin{array}{cc}
      (E\sb{1} + \Delta)\chi\sb{2} + a\sb{22} & 2\sigma a\sb{21}\\
      2\sigma a\sb{21}\sp{*} &
        E\sb{1}\chi\sb{1} + a\sb{22}
    \end{array} \right),
\label{eq:csigma}
\end{equation}
and are expressed in terms of the spin reversal invariant quantities
\begin{eqnarray}
   \chi\sb{2} &=& \langle n\sb{i\sigma}\rangle =
   \langle n\sb{i\bar\sigma}\rangle
\label{eq:chi2}\\
   \chi\sb{1} &=&
   \langle n\sb{i\sigma}\sp{h}\rangle =
   \langle n\sb{i\bar\sigma}\sp{h}\rangle
   = 1 - \chi\sb{2}
\label{eq:chi1}\\
    a\sb{22} &=& \mathcal{K}\sb{11}
             \langle\tau\sb{1}\sp{0\bar\sigma, \bar\sigma 0}\rangle -
            \mathcal{K}\sb{22}\langle\tau\sb{1}\sp{\sigma 2, 2\sigma}\rangle
\label{eq:a22}\\
   a\sb{21} &=& (\mathcal{K}\sb{11} - \mathcal{K}\sb{22})\cdot 2\sigma
                \langle\tau\sb{1}\sp{\sigma 2, \bar\sigma 0}\rangle +
  \mathcal{K}\sb{21}(\langle\tau\sb{1}\sp{0\bar\sigma, \bar\sigma 0}\rangle-
                \langle\tau\sb{1}\sp{\sigma 2, 2\sigma}\rangle).
\label{eq:a21}
\end{eqnarray}
\par
The normal hopping $2\times2$ matrix $\hat{D}\sb{ij\sigma}$ is
\emph{symmetric}, 
\begin{equation}
  \hat{D}\sb{ij\sigma}  = \left(
    \begin{array}{cc}
      d\sb{ij}\sp{22} & 2 \sigma d\sb{ij}\sp{21}\\
      2 \sigma d\sb{ij}\sp{21} & d\sb{ij}\sp{11}
    \end{array} \right)
\label{eq:Dijs}
\end{equation}
Due to the constraints~(\ref{eq:ss})--(\ref{eq:2s0s}), the charge-spin
correlations entering the matrix elements of~(\ref{eq:Dijs}) get
\emph{exactly decoupled\/} from each other, such that
\begin{eqnarray}
d\sp{22}\sb{ij} &=& \mathcal{K}\sb{22}(\chi\sb{ij}\sp{c} +
                                       \chi\sb{ij}\sp{S}) -
            \mathcal{K}\sb{11}\chi\sb{ij}\sp{s-h}
\nonumber\\
   d\sb{ij}\sp{11} &=&
\mathcal{K}\sb{11} [\chi\sb{ij}\sp{c} + (\chi\sb{1} - \chi\sb{2})\nu\sb{ij} +
                    \chi\sb{ij}\sp{S}] -
   \mathcal{K}\sb{22}\chi\sb{ij}\sp{s-h}
\nonumber\\
   d\sb{ij}\sp{21} &=& \mathcal{K}\sb{21} [(\chi\sb{ij}\sp{c} -
                       \chi\sb{2}\nu\sb{ij}) + \chi\sb{ij}\sp{S}] -
                       \mathcal{K}\sb{21} \chi\sb{ij}\sp{s-h},
\nonumber
\end{eqnarray}
with the three spin reversal invariant weighted
boson-boson correlation functions representing respectively
\emph{charge-charge} (c), \emph{spin-spin} (S), and \emph{singlet-hopping}
(s-h) correlations:
\begin{eqnarray}
      \chi\sb{ij}\sp{\rm c} &=&
      \nu\sb{ij} \langle N\sb{i}N\sb{j}\rangle / 4 ,
\label{eq:chi-c}\\
      \chi\sb{ij}\sp{\rm S} &=&
      \nu\sb{ij} \langle {\bf S}\sb{i}{\bf S}\sb{j}\rangle
\label{eq:chi-S}\\
  \chi\sb{ij}\sp{\rm s\! -\! h} &=& \nu\sb{ij}
      \langle X\sb{i}\sp{02}X\sb{j}\sp{20}\rangle
\label{eq:chi-shop}
\end{eqnarray}
In~(\ref{eq:chi-S}), ${\bf S}\sb{i} = (S\sb{i}\sp{z}, S\sb{i}\sp{\sigma})$,
with $S\sb{i}\sp{z} = (X\sb{i}\sp{\sigma\sigma} -
      X\sb{i}\sp{\bar\sigma\bar\sigma})/2$ and
      $S\sb{i}\sp{\sigma} = X\sb{i}\sp{\sigma\bar\sigma}$.
\par
The anomalous hopping $2\times 2$ matrix $\hat{\Delta}\sb{ij\sigma}$ has
a very special form namely,
\begin{equation}
  \hat{\Delta}\sb{ij\sigma} = \left(
    \begin{array}{cc}
      - \mathcal{K}\sb{21}\cdot 2\sigma &
      \frac{1}{2}(\mathcal{K}\sb{11}+\mathcal{K}\sb{22})\\
      - \frac{1}{2}(\mathcal{K}\sb{11}+\mathcal{K}\sb{22}) &
      \mathcal{K}\sb{21}\cdot 2\sigma
    \end{array} \right) \chi\sb{ij}\sp{pair}
\label{eq:Delijs}
\end{equation}
where the spin reversal invariant weighted boson-boson \emph{pairing} (pair)
correlation function is given by
\begin{eqnarray}
  \chi\sb{ij}\sp{\rm pair} &=&
      \nu\sb{ij} \langle X\sb{i}\sp{02}N\sb{j}\rangle =
      2 \nu\sb{ij} \langle X\sb{i}\sp{02}
        (X\sb{j}\sp{\sigma\sigma} + X\sb{j}\sp{22})\rangle =
\label{eq:chi-pair}\\
       &=&
      - \nu\sb{ij} \langle N\sb{j}\sp{h} X\sb{i}\sp{02}\rangle =
      - 2 \nu\sb{ij} \langle (X\sb{j}\sp{\sigma\sigma} + X\sb{j}\sp{00})
        X\sb{i}\sp{02}\rangle .
\label{eq:chi-pair1}
\end{eqnarray}
In Eqs.~(\ref{eq:chi-pair}) and~(\ref{eq:chi-pair1}), the derivation of the
second expression from the first one
makes use of the spin reversal invariance property~(\ref{eq:02s}).
\par
To get a workable expression of the frequency matrix, approximations have to
be derived for the boson-boson statistical averages entering the two-site
hopping matrix elements.
In the next section we show that the method of reference
\cite{Pl03},
yielding the pairing correlation function
$\langle X\sb{i}\sp{02}N\sb{j} \rangle$ in terms of GMFA Green functions within
an approach able to identify and rule out \emph{exponentially small terms},
can be extended to the singlet hopping correlations
$\langle X\sb{i}\sp{02}X\sb{j}\sp{20}\rangle$ as well.
\section{Hopping processes involving singlets}
\label{sec:procsing}
The right approach to the reduction of the order of correlation of the
boson-boson statistical averages
$\langle X\sb{i}\sp{02}X\sb{j}\sp{\lambda\mu}\rangle =
\langle X\sb{j}\sp{\lambda\mu} X\sb{i}\sp{02}\rangle$
goes differently for the hole-doped and electron-doped cuprates.
\par
$\bullet$ {\sf Reduction of the correlation order for hole-doped cuprates}
\par
In these systems, the Fermi level (the zero point energy) stays in the singlet
subband. We get the estimates
$E\sb{2} \simeq - \Delta$, $E\sb{2} - \Delta \simeq -2\Delta$,
      $E\sb{2} + \Delta \simeq 0$.
With $\Delta\sim 3 eV$,
$\beta\Delta \sim 3.5\cdot 10\sp{4}T\sp{-1}$.
Therefore, at $T\lesssim 300 K$, the quantities containing the factor
${\rm e}\sp{\beta E\sb{2}}\simeq {\rm e}\sp{-\beta\Delta}\lesssim
{\rm e}\sp{-100} < 10\sp{-44}$ are \emph{negligible}.
\par
We start with the following form of the spectral theorem
\cite{Zub60}
\begin{equation}
  \langle X\sb{i}\sp{02}X\sb{j}\sp{\lambda\mu}\rangle =
  \frac{\rmi}{2\pi}\int\limits_{-\infty}^{+\infty}
  \frac{{\rmd}\omega}{1+{\rme}\sp{-\beta\omega}}
  \Big[\langle\langle X\sb{i}\sp{02}|X\sb{j}\sp{\lambda\mu}\rangle\rangle
       \sb{\omega+\rmi\varepsilon} -
  \langle\langle X\sb{i}\sp{02}|X\sb{j}\sp{\lambda\mu}\rangle\rangle
       \sb{\omega-\rmi\varepsilon}\Big],
\label{eq:SpT}
\end{equation}
written for \emph{anticommutator} retarded ($\omega+\rmi\varepsilon$),
respectively advanced ($\omega-\rmi\varepsilon$) Green functions.
Their equation of motion in the (${\bf r}, \omega$)-representation is
\begin{eqnarray}
   \fl (\omega\! -\! E\sb{2})
 \langle\langle X\sb{i}\sp{02}|X\sb{j}\sp{\lambda\mu}\rangle\rangle\sb{\omega}
\simeq
   2 \langle X\sb{i}\sp{02}X\sb{j}\sp{\lambda\mu}\rangle 
   \! +\! \mathcal{K}\sb{21}\! \! \sum\sb{\sigma}\! 2\sigma\Big[
    \langle\langle\tau\sb{1,i}\sp{0\bar\sigma, 0\sigma}|
                     X\sb{j}\sp{\lambda\mu}\rangle\rangle\sb{\omega}\! -\!
    \langle\langle\tau\sb{1,i}\sp{\sigma 2,\bar\sigma 2}|
                     X\sb{j}\sp{\lambda\mu}\rangle\rangle\sb{\omega}\Big]
\label{eq:GFro}
\end{eqnarray}
where, for the sake of simplicity, the labels
$\pm \rmi\varepsilon, \varepsilon = 0\sp{+}$, describing respectively
the retarded and the advanced Green functions have been omitted.
In Eq.~(\ref{eq:GFro}), the higher order r.h.s.~contributions coming from
the inband hopping terms have been dropped off.
Replacing~(\ref{eq:GFro}) in~(\ref{eq:SpT}), we get
\begin{eqnarray}
    \fl \langle X\sb{i}\sp{02} X\sb{j}\sp{\lambda\mu} \rangle \simeq
    \mathcal{K}\sb{21}\sum\sb{\sigma}\! 2\sigma
     \int\sp{+\infty}\sb{-\infty}
   \frac{\rmd\omega}{1 + {\rme}\sp{-\beta\omega}}\times
\nonumber \\
    \times\Big(\! \! -\! \frac{1}{\pi}\Big) {\rm Im}
   \Big[ \frac{1}{\omega - E\sb{2} + \rmi\varepsilon}
   \Big(\! \langle\langle\tau\sb{1,i}\sp{0\bar\sigma ,0\sigma}|
           X\sb{j}\sp{\lambda\mu}\rangle\rangle\sb{\omega+\rmi\varepsilon}
  \! -\! \langle \langle\tau\sb{1,i}\sp{\sigma 2,\bar\sigma 2}|
           X\sb{j}\sp{\lambda\mu}\rangle\rangle\sb{\omega+\rmi\varepsilon}
   \Big)\! \Big] .
\nonumber
\end{eqnarray}
To evaluate the imaginary part, we use the identity
\cite{Zub60}
$$
  \frac{1}{\omega - E\sb{2} + \rmi\varepsilon} =
      \mathcal{P}\frac{1}{\omega - E\sb{2}} - \rmi\pi\delta(\omega - E\sb{2}).
$$
The integrals over the $\delta$-function yield (finite) GF real parts at
$\omega=E\sb{2}$, multiplied by a thermodynamic factor
$\sim {\rme}\sp{-\beta\Delta}\ll 1$.
The imaginary part of the hole subband GF
$\langle\langle\tau\sb{1,i}\sp{0\bar\sigma ,0\sigma}|
           X\sb{j}\sp{\lambda\mu}\rangle\rangle\sb{\omega + \rmi\varepsilon}$
shows a $\delta$-like maximum at $\omega=E\sb{2}-\Delta$, where
$(\omega-E\sb{2})\sp{-1}\simeq \Delta\sp{-1}$ and the thermodynamic
factor reaches a value $\sim {\rme}\sp{-2\Delta}$.
The only non-negligible contribution to the principal part integral comes
from the singlet subband GF
$\langle\langle\tau\sb{1,i}\sp{\sigma 2,\bar\sigma 2}|
       X\sb{j}\sp{\lambda\mu}\rangle\rangle\sb{\omega + \rmi\varepsilon}$
the imaginary part of which shows a $\delta$-like maximum at
$\omega=E\sb{2}+\Delta\simeq 0$.
This allows us to approximate
$(\omega-E\sb{2})\sp{-1}\approx \Delta\sp{-1}$
within the integral over the singlet subband GF to get
\begin{equation}
  \langle X\sb{i}\sp{02}X\sb{j}\sp{\lambda\mu}\rangle \simeq
  (1-\delta\sb{ij})\frac{\mathcal{K}\sb{21}}{\Delta}
  \sum\sb{\sigma}2\bar\sigma
  \langle\tau\sb{1,i}\sp{\sigma 2,\bar\sigma 2}
       X\sb{j}\sp{\lambda\mu}\rangle
\label{eq:red2-s}
\end{equation}
Replacing this result in Eq.~(\ref{eq:chi-pair}) and using~(\ref{eq:taup})
we get
\begin{equation}
\fl \qquad \quad \chi\sb{ij}\sp{pair} \simeq
  (1-\delta\sb{ij})\frac{\mathcal{K}\sb{21}\nu\sb{ij}}{\Delta}
  \Big[ 4 \nu\sb{ij}\! \cdot\!
  2\bar\sigma\langle X\sb{i}\sp{\sigma 2} X\sb{j}\sp{\bar\sigma 2}\rangle
   \! -\! \! \! \sum\sb{m\neq (i,j)}\! \!
  \nu\sb{im}\! \sum\sb{\sigma} 2\sigma
  \langle X\sb{i}\sp{\sigma 2}X\sb{m}\sp{\bar\sigma 2}N\sb{j}\rangle
  \Big]
\label{eq:p-hd-2s3s}
\end{equation}
Omitting the three-site terms, we get the two-site approximation of the
superconducting pairing originating in the singlet subband,
\begin{equation}
  \chi\sb{ij}\sp{pair} \simeq 
  (1-\delta\sb{ij})\frac{4 \mathcal{K}\sb{21}\nu\sb{ij}\sp{2}}{\Delta}\cdot
  2\bar\sigma\langle X\sb{i}\sp{\sigma 2} X\sb{j}\sp{\bar\sigma 2}\rangle ,
\label{eq:p-hd-fin0}
\end{equation}
which reproduces the well-known two-site exchange term of the $t$-$J$ model.
\par
For the singlet hopping correlation function, (\ref{eq:red2-s}) yields
the two-site approximation
\begin{equation}
  \chi\sb{ij}\sp{s-h} \simeq 
  (1-\delta\sb{ij})\frac{2 \mathcal{K}\sb{21}\nu\sb{ij}\sp{2}}{\Delta}\cdot
  2\bar\sigma\langle X\sb{i}\sp{\sigma 2}X\sb{j}\sp{\bar\sigma 0}\rangle
\label{eq:sh-hd-fin0}
\end{equation}
\par
$\bullet$ {\sf Reduction of the correlation order for electron-doped cuprates}
\par
The Fermi level (the zero point energy) stays now in the hole subband.
We have the estimates
$E\sb{2} \simeq \Delta$, $E\sb{2} + \Delta \simeq 2\Delta$,
      $E\sb{2} - \Delta \simeq 0$.
\par
It is convenient now to start with
the alternative form of the spectral theorem
\cite{Zub60}
\begin{equation}
  \langle X\sb{j}\sp{\lambda\mu}X\sb{i}\sp{02}\rangle =
  \frac{\rmi}{2\pi}\int\limits_{-\infty}^{+\infty}
  \frac{{\rmd}\omega}{{\rm e}\sp{\beta\omega}+1}
  \Big[\langle\langle X\sb{i}\sp{02}|X\sb{j}\sp{\lambda\mu}\rangle\rangle
       \sb{\omega+\rmi\varepsilon} -
  \langle\langle X\sb{i}\sp{02}|X\sb{j}\sp{\lambda\mu}\rangle\rangle
       \sb{\omega-\rmi\varepsilon}\Big],
\label{eq:SpTa}
\end{equation}
with the retarded and advanced GFs following from the same
equation~(\ref{eq:GFro}).
\par
Exponentially small quantities result from the $\delta$-term of
$(\omega-E\sb{2}+\rmi\varepsilon)\sp{-1}$
and from the singlet subband GF
$\langle\langle\tau\sb{1,i}\sp{\sigma 2,\bar\sigma 2}|
       X\sb{j}\sp{\lambda\mu}\rangle\rangle\sb{\omega + \rmi\varepsilon}$.
The hole subband GF
$\langle\langle\tau\sb{1,i}\sp{0\bar\sigma ,0\sigma}|
           X\sb{j}\sp{\lambda\mu}\rangle\rangle\sb{\omega + \rmi\varepsilon}$,
yields the non-negligible contribution
\begin{equation}
   \langle X\sb{j}\sp{\lambda\mu}X\sb{i}\sp{02}\rangle \simeq
  (1-\delta\sb{ij})\frac{\mathcal{K}\sb{21}}{\Delta}
  \sum\sb{\sigma}2\bar\sigma
  \langle X\sb{j}\sp{\lambda\mu}\tau\sb{1,i}\sp{0\bar\sigma ,0\sigma }\rangle
\label{eq:red2-e}
\end{equation}
Replacing in~(\ref{eq:chi-pair1}) and
omitting the three-site terms, we get the two-site approximation of the
superconducting pairing originating in the hole subband,
\begin{equation}
  \chi\sb{ij}\sp{pair} \simeq
  (1-\delta\sb{ij})\frac{4 \mathcal{K}\sb{21}\nu\sb{ij}\sp{2}}{\Delta}\cdot
   2\sigma\langle X\sb{i}\sp{0\bar\sigma} X\sb{j}\sp{0\sigma}\rangle
\label{eq:p-ed-fin0}
\end{equation}
Finally, the two-site approximation of the singlet-hopping correlation
function is
\begin{equation}
   \langle X\sb{i}\sp{02}X\sb{j}\sp{20}\rangle \simeq
  (1-\delta\sb{ij})\frac{2 \mathcal{K}\sb{21}\nu\sb{ij}\sp{2}}{\Delta}\cdot
   2\bar\sigma\langle X\sb{i}\sp{0\bar\sigma} X\sb{j}\sp{2\sigma}\rangle .
\label{eq:sh-ed-fin0}
\end{equation}
\par
In conclusion, the GMFA superconducting pairing is a second order effect.
The lowest order contribution to it originates in interband hopping
correlating annihilation (or creation) of pairs of spins at neighbouring
lattice sites $i$ and $j$ within that energy subband which crosses the Fermi
level.
\par
Similarly, the singlet hopping is a second order effect as well.
It mainly proceeds by interband $i\rightleftarrows j$ single particle jumps
from the upper energy subband to the lower energy subband.
\section{Frequency matrix in (${\bf q}, \omega$)-representation}
\label{sec:freqmatqo}
The calculation of the matrix elements of
$\tilde \mathcal{A}\sb{\sigma}({\bf q})$ from Eq.~(\ref{eq:Aq2r}) asks for
three essentially different kinds of Fourier transforms, namely,
\begin{itemize}
  \item
    The averages of the Hubbard $1$-forms entering Eqs.~(\ref{eq:a22})
    and~(\ref{eq:a21}) result in sums of products of ${\bf q}$-space averages
    and geometrical form factors:
    \begin{equation}
       \langle\tau\sb{1,i}\sp{\lambda\mu, \nu\varphi}\rangle =
       \sum\sb{\alpha=1}\sp{3}\nu\sb{\alpha}\cdot
       \frac{1}{N}\sum\sb{\bf q}\langle X\sp{\lambda\mu}
         X\sp{\nu\varphi}\rangle\sb{\bf q}\gamma\sb{\alpha}({\bf q})
    \label{eq:tau1o}
    \end{equation}
    for label sets $\{(\lambda\mu, \nu\varphi)\} \in
    \{(0\bar\sigma, \bar\sigma 0); (\sigma 2, 2\sigma);
      (\sigma 2, \bar\sigma 0)\}$.
    \par
    The quantity $\langle X\sp{\lambda\mu} X\sp{\nu\varphi}\rangle\sb{\bf q}$
    denotes the average of the ${\bf q}$-space image of the product of
    Hubbard operators of labels $\lambda\mu$ and $\nu\varphi$ respectively,
     \begin{equation}
       \fl \qquad \langle X\sp{\lambda\mu}X\sp{\nu\varphi}\rangle\sb{\bf q} =
        \frac{\rmi}{2\pi}\int\limits_{-\infty}^{+\infty}
        \frac{{\rmd}\omega}{1+{\rme}\sp{-\beta\omega}}
        \Big[G\sp{\lambda\mu, \nu\varphi}({\bf q}, \omega+\rmi\varepsilon) -
        G\sp{\lambda\mu, \nu\varphi}({\bf q}, \omega-\rmi\varepsilon)\Big]
     \label{eq:SpTq}
     \end{equation}
     Finally, in Eq.~(\ref{eq:tau1o}), $\gamma\sb{\alpha}({\bf q})$ denote the
     nn $(\alpha=1)$, nnn $(\alpha=2)$, and third neighbour $(\alpha=3)$
     geometrical form factors,
   $\gamma\sb{1}({\bf q}) = 2 [\cos(q\sb{x}a\sb{x}) + \cos(q\sb{y}a\sb{y})]$,
   $\gamma\sb{2}({\bf q}) = 4 \cos(q\sb{x}a\sb{x}) \cos(q\sb{y}a\sb{y})$,
   $\gamma\sb{3}({\bf q}) = 2 [\cos(2q\sb{x}a\sb{x}) + \cos(2q\sb{y}a\sb{y})]$.
  \item
     For the two-site weighted singlet hopping~(\ref{eq:chi-shop})
     and the superconducting pairing~(\ref{eq:chi-pair}), the Fourier
     transforms result in \emph{convolutions\/} of specific averages and
     geometrical form factors. The results are as follows:
     \par
     {\bf $-$} \emph{Singlet hopping}
    \begin{equation}
       \chi\sp{\rm s-h} ({\bf q}) =
       \sum\sb{\alpha=1}\sp{3}\nu\sb{\alpha}\sp{2}\cdot
       \frac{1}{N}\sum\sb{\bf k}\Xi\sb{\bf k}
       \gamma\sb{\alpha}({\bf q}-{\bf k})
    \label{eq:shhd1}
    \end{equation}
    where $\Xi\sb{\bf k} = 2\sigma\langle X\sp{\sigma 2}
         X\sp{\bar\sigma 0}\rangle\sb{\bf k}$,
    while $\Xi\sb{\bf k} = 2\sigma\langle X\sp{0\bar\sigma}
         X\sp{2\sigma}\rangle\sb{\bf k}$ for hole-doped and
    electron-doped cuprates respectively, with averages defined
    in~(\ref{eq:SpTq}).
     \par
     {\bf $-$} \emph{Superconducting pairing}
    \begin{equation}
       \chi\sp{\rm pair} ({\bf q}) =
       \sum\sb{\alpha=1}\sp{3}\nu\sb{\alpha}\sp{2}\cdot
       \frac{1}{N}\sum\sb{\bf k}\Pi\sb{\bf k}
       \gamma\sb{\alpha}({\bf q}-{\bf k})
    \label{eq:phd1}
    \end{equation}
    where $\Pi\sb{\bf k} = 2\bar\sigma\langle X\sp{\sigma 2}
         X\sp{\bar\sigma 2}\rangle\sb{\bf k}$,
    while $\Pi\sb{\bf k} = 2\sigma\langle X\sp{0 \bar\sigma}
         X\sp{0\sigma}\rangle\sb{\bf k}$  for hole-doped and
    electron-doped cuprates respectively, with averages defined
    in~(\ref{eq:SpTq}).
  \item
    The charge-charge and spin-spin correlation functions~(\ref{eq:chi-c})
    and~(\ref{eq:chi-S}) are treated approximately following
    \cite{Pl95,Pl06}:
    \par
     -- The order of the \emph{charge-charge} correlation
     function $\langle N\sb{i} N\sb{j} \rangle$ is lowered using a
     Hubbard type~I approximation decoupling procedure
     $\langle N\sb{i} N\sb{j} \rangle \simeq \langle N\sb{i}\rangle
                   \langle N\sb{j}\rangle = 2 \chi\sb{2}$.
    \par
     -- The \emph{spin-spin} correlation function
      $\langle {\bf S}\sb{i}{\bf S}\sb{j}\rangle$
      is kept undecoupled, but treated phenomenologically.
      Eq.~(\ref{eq:taup}) implies the occurrence of up to three non-vanishing
      spin-spin correlation functions: nn,
      $\chi\sb{1}\sp{S} = \langle S\sb{i} S\sb{i\pm a\sb{x/y}}\rangle$,
      nnn,
  $\chi\sb{2}\sp{S} = \langle S\sb{i} S\sb{i \pm a\sb{x} \pm a\sb{y}}\rangle$,
      and
      $\chi\sb{3}\sp{S} = \langle S\sb{i} S\sb{i\pm 2a\sb{x/y}}\rangle$.
      These are site independent quantities.
\end{itemize}
Using the above results, we get from~(\ref{eq:Aq2r}) and~(\ref{eq:Aijs})
the mathematical structure of the frequency matrix
$\tilde \mathcal{A}\sb{\sigma}({\bf q})$ as follows,
\begin{equation}
  \tilde \mathcal{A}\sb{\sigma}({\bf q}) =
  \left( \begin{array}{cc}
    \hat E\sb{\sigma}({\bf q})& \hat \Phi\sb{\sigma}({\bf q})\\
    (\hat \Phi\sb{\sigma}({\bf q}))\sp{\dagger} &
      - (\hat {E}\sb{\bar\sigma}({\bf q}))\sp{\top}
  \end{array} \right) .
\label{eq:Aqs}
\end{equation}
The \emph{normal\/} $2\times 2$ matrix contributions to
$\tilde \mathcal{A}\sb{\sigma}({\bf q})$ show the characteristic
$\sigma$-dependence,
\begin{equation}
  \hat E\sb{\sigma}({\bf q}) =
  \left( \begin{array}{cc}
    c\sb{22} & 2\sigma c\sb{21}\\
    2\sigma c\sb{21}\sp{*}&c\sb{11}
  \end{array} \right) ; \quad
  - (\hat {E}\sb{\bar\sigma}({\bf q}))\sp{\top} =
  \left( \begin{array}{cc}
    - c\sb{22} & 2\sigma c\sb{21}\sp{*}\\
    2\sigma c\sb{21}& -c\sb{11}
  \end{array} \right)
\label{eq:Eqs}
\end{equation}
with the $\sigma$-independent terms $c\sb{ab}$ carrying normal one-site
and two-site matrix elements,
\begin{eqnarray}
   c\sb{22} &\equiv&c\sb{22}({\bf q}) = (E\sb{1}+\Delta)\chi\sb{2} +
             a\sb{22} + d\sb{22}({\bf q})
\nonumber\\
   c\sb{11} &\equiv&c\sb{11}({\bf q}) = E\sb{1}\chi\sb{1} +
             a\sb{22} + d\sb{11}({\bf q})
\nonumber\\
   c\sb{21} &\equiv&c\sb{21}({\bf q}) = a\sb{21} + d\sb{21}({\bf q})
\nonumber\\
   d\sb{ab}({\bf q}) &=& \mathcal{K}\sb{ab}\sum\sb{\alpha=1}\sp{3}
      \nu\sb{\alpha}\gamma\sb{\alpha}({\bf q})[\chi\sb{\alpha}\sp{S} +
      (-1)\sp{a+b}\chi\sb{a}\chi\sb{b}] +
      \frac{1}{2} J\sb{ab} \chi\sp{\rm s-h}({\bf q})
\nonumber
\end{eqnarray}
The one-site terms are defined by Eqs.~(\ref{eq:a22})--(\ref{eq:a21})
and~(\ref{eq:tau1o}). The exchange energy parameters are given by
\begin{equation}
     J\sb{ab} = 4\mathcal{K}\sb{ab} \mathcal{K}\sb{21}/\Delta , \quad
          \{ab\} \in \{22, 11, 21\} ,
\label{eq:Jab}
\end{equation}
while the singlet hopping contribution $\chi\sp{\rm s-h}({\bf q})$ is given by
Eq.~(\ref{eq:shhd1}).
\par
The \emph{anomalous\/} $2\times 2$ matrix contributions to
$\tilde \mathcal{A}\sb{\sigma}({\bf q})$, obtained from~(\ref{eq:Delijs}),
show the characteristic $\sigma$-dependence,
\begin{equation}
  \hat \Phi\sb{\sigma}({\bf q}) =
  \left( \begin{array}{cc}
    - 2\sigma\xi\sb{1} b & \xi\sb{2}b\\
    - \xi\sb{2}b& 2\sigma\xi\sb{1} b
  \end{array} \right) ; \quad
  (\hat \Phi\sb{\sigma}({\bf q}))\sp{\dagger} =
  \left( \begin{array}{cc}
    - 2\sigma \xi\sb{1} b\sp{*} & - \xi\sb{2}b\sp{*}\\
    \xi\sb{2}b\sp{*} & 2\sigma\xi\sb{1} b\sp{*}
  \end{array} \right)
\label{eq:Fqs}
\end{equation}
with $\xi\sb{1} = J\sb{21}$, $\xi\sb{2} = (J\sb{11} + J\sb{22})/2$,
whereas $b\equiv b({\bf q})$ is a shorthand notation for the pairing matrix
element~(\ref{eq:phd1}).
\begin{rem}
\label{rem:GF}
  The spin reversal
  $\sigma \rightarrow \bar\sigma$ symmetry properties of the elemental Green
  functions entering the matrix GF~(\ref{eq:GF}) are \emph{identical} to those
  established for the underlying frequency matrix
  $\tilde \mathcal{A}\sb{\sigma}({\bf q})$.
\end{rem}
\section{GMFA Green function}
\label{sec:GMFA-GF}
From Eqs.~(\ref{eq:avss}) and~(\ref{eq:av02}) it follows that the matrix
$\tilde\chi$, Eq.~(\ref{eq:chi}), is diagonal and spin reversal invariant,
with two nonvanishing matrix elements,
\begin{eqnarray}
  \tilde\chi &=& \left(
    \begin{array}{cc}
      \hat{\chi} & \hat{0}\\
         \hat{0} & \hat{\chi}
    \end{array}
              \right), \quad
  \hat{\chi} = \left(
    \begin{array}{cc}
      \chi\sb{2} & 0\\
               0 & \chi\sb{1}
    \end{array}
              \right), \quad
  \hat{0} = \left(
    \begin{array}{cc}
      0 & 0\\
      0 & 0
    \end{array}
              \right) ,
\label{eq:chicalc}
\end{eqnarray}
where $\chi\sb{2}$ and $\chi\sb{1}$ are given by Eqs.~(\ref{eq:chi2})
and~(\ref{eq:chi1}) respectively.
\par
Replacing in~(\ref{eq:gf0}) the expressions~(\ref{eq:chicalc}) of the matrix
$\tilde\chi$ and~(\ref{eq:Aqs}) of the frequency matrix
$\tilde \mathcal{A}\sb{\sigma}({\bf q})$, we get a structure of the GMFA-GF
matrix obeying the general symmetry properties established in
\cite{Pl97},
\begin{equation}
  \tilde G\sp{0}\sb{\sigma}({\bf q}, \omega) =
  \left( \begin{array}{cc}
    \hat G\sp{0}\sb{\sigma}({\bf q}, \omega)&
    \hat F\sp{0}\sb{\sigma}({\bf q}, \omega)\\
    (\hat F\sp{0}\sb{\sigma}({\bf q}, \omega))\sp{\dagger} &
      - (\hat G\sp{0}\sb{\bar\sigma}({\bf q}, -\omega))\sp{\top}
  \end{array} \right) ,
\label{eq:Gtldqs}
\end{equation}
where the argument $\omega$ carries, in fact, the complex
value $\omega + i\varepsilon, \varepsilon = 0\sp{+}$. (Hence the elemental
GFs containing the argument $\omega$ point to \emph{retarded\/} GFs, while
those containing the argument $-\omega$ point to \emph{advanced\/} GFs.)
\par
The \emph{normal\/} $2\times 2$ matrix
$\hat G\sp{0}\sb{\sigma}({\bf q}, \omega)$ shows the characteristic
$\sigma$-dependence,
\begin{equation}
  \hat G\sp{0}\sb{\sigma}({\bf q}, \omega) =
  \left( \begin{array}{cc}
    g\sb{22}({\bf q},\omega) & 2\sigma g\sb{21}({\bf q},\omega)\\
    2\sigma g\sb{21}\sp{*}({\bf q},\omega) & g\sb{11}({\bf q},\omega)
  \end{array} \right)\cdot
    \frac{1}{\mathcal{D}({\bf q}, \omega)}
\label{eq:Ghatqs}
\end{equation}
with the $\sigma$-independent components $g\sb{ab}({\bf q},\omega)$ found
from
$$
   g\sb{ab}({\bf q},\omega) = A\sb{ab}\omega\sp{3} + B\sb{ab}\omega\sp{2} +
       C\sb{ab}\omega + D\sb{ab}, \quad \{ab\}\in\{22,11,21\}.
$$
Here the coefficients $A\sb{ab}$ are given respectively by
$$
    A\sb{22} = \chi\sb{2}, \quad A\sb{11} = \chi\sb{1}, \quad A\sb{21} = 0,
$$
while $B\sb{ab}$, $C\sb{ab}$, $D\sb{ab}$ are ${\bf q}$-dependent coefficients:
$$
   B\sb{22}({\bf q}) = c\sb{22}, \quad 
   B\sb{11}({\bf q}) = c\sb{11}, \quad 
   B\sb{21}({\bf q}) = c\sb{21}
$$
\begin{eqnarray}
  C\sb{22}({\bf q})&=&-[\chi\sb{2}(c\sb{11}\sp{2} + \xi\sb{1}\sp{2}|b|\sp{2})+
                        \chi\sb{1}(|c\sb{21}|\sp{2} + \xi\sb{2}\sp{2}|b|\sp{2})
                        ]/\chi\sb{1}\sp{2}
\nonumber\\
   C\sb{11}({\bf q})&=&-[\chi\sb{1}(c\sb{22}\sp{2} + \xi\sb{1}\sp{2}|b|\sp{2})+
                        \chi\sb{2}(|c\sb{21}|\sp{2} + \xi\sb{2}\sp{2}|b|\sp{2})
                        ]/\chi\sb{2}\sp{2}
\nonumber\\
    C\sb{21}({\bf q})&=&[c\sb{21}(\chi\sb{2}c\sb{11} + \chi\sb{1}c\sb{22}) -
                        \xi\sb{1}\xi\sb{2}|b|\sp{2}] / (\chi\sb{1}\chi\sb{2})
\nonumber
\end{eqnarray}
\begin{eqnarray}
\fl \qquad \quad D\sb{22}({\bf q}) =
   -[c\sb{11}(c\sb{22}c\sb{11} - |c\sb{21}|\sp{2}) +
    (c\sb{22} \xi\sb{1}\sp{2} + c\sb{11}\xi\sb{2}\sp{2} +
                         2 \Re (c\sb{21})\xi\sb{1}\xi\sb{2})|b|\sp{2}
                        ]/\chi\sb{1}\sp{2}
\nonumber\\
\fl \qquad \quad D\sb{11}({\bf q}) =
     -[c\sb{22}(c\sb{22}c\sb{11} - |c\sb{21}|\sp{2}) +
     (c\sb{11} \xi\sb{1}\sp{2} + c\sb{22}\xi\sb{2}\sp{2} +
                         2 \Re (c\sb{21})\xi\sb{1}\xi\sb{2})|b|\sp{2}
                        ]/\chi\sb{2}\sp{2}
\nonumber\\
\fl \qquad \quad D\sb{21}({\bf q}) =
   \{c\sb{21}(c\sb{22}c\sb{11} - |c\sb{21}|\sp{2}) -
    [c\sb{21}\sp{*}\xi\sb{1}\sp{2} + c\sb{21}\xi\sb{2}\sp{2} +
    (c\sb{22} + c\sb{11})\xi\sb{1}\xi\sb{2}]|b|\sp{2}
                        \} / (\chi\sb{1}\chi\sb{2})
\nonumber
\end{eqnarray}
\par
The \emph{anomalous\/} $2\times 2$ matrix
$\hat F\sp{0}\sb{\sigma}({\bf q}, \omega)$ shows the characteristic
$\sigma$-dependence,
\begin{equation}
  \hat F\sp{0}\sb{\sigma}({\bf q}, \omega) =
  \left( \begin{array}{cc}
    2\sigma f\sb{22}({\bf q},\omega) & f\sb{21}({\bf q},\omega)\\
    - f\sb{21}({\bf q},- \omega) & 2\sigma f\sb{11}({\bf q},\omega)
  \end{array} \right)\cdot
    \frac{1}{\mathcal{D}({\bf q}, \omega)}
\label{eq:Fhatqs}
\end{equation}
with the elemental GFs $f\sb{ab}({\bf q}, \omega)$ given by
\begin{eqnarray}
   f\sb{aa}({\bf q},\omega) &=& (P\sb{aa}\omega\sp{2} +
     R\sb{aa})b, \quad \{aa\}\in\{22,11\},
\nonumber\\
   f\sb{21}({\bf q},\omega) &=& (P\sb{21}\omega\sp{2} + Q\sb{21}\omega +
     R\sb{21})b.
\nonumber
\end{eqnarray}
Here, $P\sb{22} = -\xi\sb{1}$, $P\sb{11} = \xi\sb{1}$, and
$P\sb{21} = -\xi\sb{2}$ are ${\bf q}$-independent, while
\begin{eqnarray}
   R\sb{22}({\bf q}) =
    [(c\sb{11}\sp{2}+c\sb{21}\sp{2})\xi\sb{1} +
     2 c\sb{11}c\sb{21}\xi\sb{2} + \xi\sb{1}(\xi\sb{1}\sp{2} -\xi\sb{2}\sp{2})
                         |b|\sp{2}]/\chi\sb{1}\sp{2}
\nonumber\\
   R\sb{11}({\bf q}) =
     -[(c\sb{22}\sp{2} + {c\sb{21}\sp{*}}\sp{2})\xi\sb{1} +
   2c\sb{22}c\sb{21}\sp{*}\xi\sb{2}+\xi\sb{1}(\xi\sb{1}\sp{2}-\xi\sb{2}\sp{2}) 
                        |b|\sp{2}]/\chi\sb{2}\sp{2}
\nonumber\\
   R\sb{21}({\bf q}) =
    [(c\sb{11}c\sb{21}\sp{*} + c\sb{22}c\sb{21})\xi\sb{1} +
    (c\sb{22}c\sb{11} + |c\sb{21}|\sp{2})\xi\sb{2} -
      \xi\sb{2}(\xi\sb{1}\sp{2} - \xi\sb{2}\sp{2})
                  |b|\sp{2}] / (\chi\sb{1}\chi\sb{2})
\nonumber\\
    Q\sb{21}({\bf q}) =
     [(\chi\sb{2}c\sb{21} - \chi\sb{1}c\sb{21}\sp{*})\xi\sb{1} +
      (\chi\sb{2}c\sb{11} - \chi\sb{1}c\sb{22})\xi\sb{2}
                  ] / (\chi\sb{1}\chi\sb{2}).
\nonumber
\end{eqnarray}
The denominator $\mathcal{D}({\bf q}, \omega)$ occurring in
Eqs.~(\ref{eq:Ghatqs}) and~(\ref{eq:Fhatqs}), which is proportional to
the determinant of the matrix
$\tilde{\chi}\omega - \tilde{A}\sb{\sigma}({\bf q})$ in~(\ref{eq:gf0}),
shows the following monic bi-quadratic dependence in $\omega$:
\begin{equation}
   \mathcal{D}({\bf q}, \omega) = (\omega\sp{2} - u\omega + v)
                                  (\omega\sp{2} + u\omega + v),
\label{eq:Delta}
\end{equation}
where $v=v({\bf q})$ and $u=u({\bf q})$ are found respectively from
\begin{eqnarray}
   v\sp{2} =
   \Big\{\Big[(c\sb{22}c\sb{11}\! -\! |c\sb{21}|\sp{2}) -
   (\xi\sb{1}\sp{2}-\xi\sb{2}\sp{2})|b|\sp{2}\Big]\sp{2} +
    \Big[ [(c\sb{22}\! +\! c\sb{11}) + 2\Re (c\sb{21})]\sp{2}
      \xi\sb{1}\sp{2} -
\nonumber\\
     \qquad - 4 (c\sb{22}\! +\! c\sb{11})\Re (c\sb{21})
      \xi\sb{1}(\xi\sb{1} - \xi\sb{2}) -
     4 |c\sb{21}|\sp{2}(\xi\sb{1}\sp{2}\! -\! \xi\sb{2}\sp{2})\Big]
                  |b|\sp{2}\Big\} / (\chi\sb{1}\sp{2}\chi\sb{2}\sp{2})
\label{eq:v2}\\
   u\sp{2} - 2 v =
      \frac{1}{\chi\sb{1}\sp{2}}(c\sb{11}\sp{2}+\xi\sb{1}\sp{2}|b|\sp{2}) +
      \frac{1}{\chi\sb{2}\sp{2}}(c\sb{22}\sp{2}+\xi\sb{1}\sp{2}|b|\sp{2}) +
     \frac{2}{\chi\sb{1}\chi\sb{2}}(|c\sb{21}|\sp{2}+\xi\sb{2}\sp{2}|b|\sp{2}).
\label{eq:u2}
\end{eqnarray}
  A necessary consistency condition to be satisfied by the parameters of the
model at any vector ${\bf q}$ inside the Brillouin zone is
$v\sp{2}({\bf q}) \ge 0$.
\begin{rem}
The zeros of the determinant of the GMFA-GF, 
\begin{equation}
   \mathcal{D}({\bf q}, \omega) = 0
\label{eq:zeroDelta}
\end{equation}
provide the GMFA energy spectrum of the system.
\par
At every wave vector ${\bf q}$ inside the Brillouin zone, this yields for
the superconducting state the energy eigenvalue set
\begin{eqnarray}
  && \{ \Omega\sb{1}({\bf q}), \ \Omega\sb{2}({\bf q}), \
     - \Omega\sb{2}({\bf q}), \ - \Omega\sb{1}({\bf q}) \},
\nonumber\\
  && \Omega\sb{1,2}({\bf q}) = (u/2)\pm \sqrt{(u/2)\sp{2} - v}.
\label{eq:Om12}
\end{eqnarray}
\end{rem}
\par
In the normal state $(b=0)$, Eqs.~(\ref{eq:v2}) and (\ref{eq:u2}) reduce
respectively to
\begin{eqnarray}
   v\sb{0} &=& (c\sb{22} / \chi\sb{2}) (c\sb{11} / \chi\sb{1}) -
         |c\sb{21}|\sp{2} / (\chi\sb{1}\chi\sb{2})
\nonumber\\
   u\sb{0} &=& (c\sb{22} / \chi\sb{2}) + (c\sb{11} / \chi\sb{1})
\nonumber
\end{eqnarray}
such that the energy spectrum is given by the roots of the second order
equation $\omega\sp{2} - u\sb{0}\omega + v\sb{0} = 0$ solved in
{\rm \cite{Pl95}}.
\par
Finally, if we assume a pure Hubbard model (i.e., \emph{energy band
independent hopping parameters},
$\mathcal{K}\sb{11} = \mathcal{K}\sb{22} = \mathcal{K}\sb{21} \equiv t$,
\cite{Pl06}),
then a significant simplification of the equations derived in the last two
sections is obtained.
The normal $2\times 2$ matrix $\hat E\sb{\sigma}({\bf q})$ becomes
\emph{symmetric\/} and so is the normal GMFA-GF
$\hat G\sp{0}\sb{\sigma}({\bf q}, \omega)$.
Moreover, there is a single exchange energy parameter in~(\ref{eq:Fqs}),
$\xi\sb{1} = \xi\sb{2} \equiv J = 4 t\sp{2}/\Delta$, which
simplifies the anomalous $2\times 2$ frequency matrix to
$\hat \Phi\sb{\sigma}({\bf q}) =
  \left( \begin{array}{cc}
    2\sigma & 1 \\
    - 1     & 2\sigma
  \end{array} \right) J b $,
such that the quantities $u$ and $v$ in the expression~(\ref{eq:Delta}) of
the GF determinant reduce to
\begin{eqnarray}
   v\sp{2} =
   \big[(c\sb{22}c\sb{11} - c\sb{21}\sp{2})\sp{2} +
    (c\sb{22} + c\sb{11} + 2c\sb{21})\sp{2} J\sp{2}
                  |b|\sp{2}\big] / (\chi\sb{1}\sp{2}\chi\sb{2}\sp{2})
\label{eq:v2t}\\
   u\sp{2} - 2 v =
     \big[ \chi\sb{2}\sp{2} c\sb{11}\sp{2} + \chi\sb{1}\sp{2}c\sb{22}\sp{2} +
     2\chi\sb{1}\chi\sb{2}c\sb{21}\sp{2}+J\sp{2}|b|\sp{2}
     \big] / (\chi\sb{1}\sp{2}\chi\sb{2}\sp{2}).
\label{eq:u2t}
\end{eqnarray}
A non-negative value $v \geq 0$ always follows from Eq.~(\ref{eq:v2t}),
however, the reality of the solutions~(\ref{eq:Om12}) needs investigation
of the domain of variation of the adjustable parameters of the model.
\section{Conclusions}
\label{sec:concl}
The two-band Hubbard model of the high $T\sb{c}$ superconductivity in cuprates
\cite{Pl95,Pl03}
uses Hubbard operator algebra on a physical system characterized by specific
invariance symmetries with respect to translations and spin reversal.
\par
In the present paper we have shown that the system symmetries result either
in invariance properties or exact vanishing of several characteristic
statistical averages.
The vanishing of the one-site anomalous matrix elements
is shown to be a property which is embedded in the Hubbard operator algebra.
Another worth mentioning consequence following from the spin reversal
invariance properties of the two-site statistical averages is the \emph{exact
decoupling\/} from each other of the charge and spin correlations entering
the matrix elements of the frequency matrix.
The use of these results allowed rigorous derivation and simplification
of the expression of the frequency matrix of the generalized mean field
approximation (GMFA) Green function (GF) matrix of the model.
\par
For the higher order boson-boson
averages $\langle X\sb{i}\sp{02}X\sb{j}\sp{20}\rangle$ and
$\langle X\sb{i}\sp{02}N\sb{j}\rangle$, which enter respectively the normal
singlet hopping and anomalous exchange pairing contributions to the
frequency matrix, an approximation procedure resulting in GMFA-GF expressions
was described.
The procedure avoids the current decoupling schemes
\cite{Roth,Mancini}.
Its principle, first formulated in
\cite{Pl03},
consists in the identification and elimination of exponentially small
contributions to the spectral theorem representations of these statistical
averages.
\par
A point worth noting is that the proper identification of exponentially small
quantities asks for the use of different starting expressions of the spectral
theorem for the hole-doped and electron-doped cuprates.
\par
The results of the reduction procedure may be summarized as follows:
\begin{itemize}
   \item
     The singlet hopping is a second order effect which may be described as
     interband $i\rightleftarrows j$ single particle jumps from the upper
     to the lower energy subband.
   \item
     The GMFA superconducting pairing is a second order effect, the lowest
     order contribution to which originates in interband hopping correlating
     the annihilation (creation) of spin pairs at neighbouring lattice
     sites $i$ and $j$ within that energy subband which crosses the Fermi
     level.
\end{itemize}
\par
The derivation of the most general and simplest possible expressions of the
frequency matrix and of the GMFA-GF matrix in the
$({\bf q}, \omega)$-representation enables reliable numerical investigation
of the consequences coming from the adjustable parameters of the model (the
degree of hole/electron doping, the energy gap $\Delta$, the hopping
parameters).
\par
Another open question of the GF approach to the solution of the present model
is the use of the Hubbard operator algebra to get
rigorous derivation and simplification of the Dyson equation of the complete
Green function. As shown previously in
\cite{Pl03},
the self-energy corrections induce a spin fluctuation $d$-wave pairing
originating in kinematic interaction in the second order.
\par
These investigations are underway and results will be reported in a
forthcoming paper.
\ack{The authors would like to express their gratitude to Prof. N.M.~Plakida
     for useful advice and critical reading of the manuscript.
     Partial financial support was secured by the Romanian Authority
     for Scientific Research (Project 11404/31.10.2005 - SIMFAP).}

\vfil\eject
\def\endpage{\hfill\eject}

\end{document}